Jean Bricmont
FYMA, 2, ch. du Cyclotron
B-1348 Louvain-la-neuve, Belgium
Tel:32-10-473277
Fax: 32-10-472414
\catcode`\@=12 {\count255=\time\divide\count255 by 60
\xdef\hourmin{\number\count255} \multiply\count255 by-60\advance\count255
by\time \xdef\hourmin{\hourmin:\ifnum\count255<10 0\fi\the\count255}}
\def\ps@draft{\let\@mkboth\@gobbletwo \def\@oddhead{} \def\@oddfoot {\hbox
to
7 cm{$\caliptstyle Draft\ version:\ \draftdate$ \hfil} \hskip
-7cm\hfil\rm\thepage \hfil} \def\@evenhead{}\let\@evenfoot\@oddfoot}
\catcode`\@=12 \def\draftdate{\number\month/\number\day/\number\year\ \ \
\hourmin } 
\documentstyle[12pt]{article} 
 \newcommand{\be}{\begin{eqnarray}}
\newcommand{\BE}{\begin{eqnarray}}
\newcommand{\en}{\end{eqnarray}}
\newcommand{\EN}{\end{eqnarray}}\newcommand{\non}{\nonumber}
\newcommand{\no}{\noindent} \newcommand{\vs}{\vspace}
  
  \newcommand{\Bbb}{\bf}
 \newcommand{\ha}{{1\over 2}}
\title{{\bf Renormalizing Partial Differential Equations}}

\author{J.Bricmont\thanks{Supported by EC grants SC1-CT91-0695 and
CHRX-CT93-0411}\\ UCL, Physique Th\'eorique, B-1348, Louvain-la-Neuve,
Belgium\\bricmont@fyma.ucl.ac.be\and A.Kupiainen\thanks{Supported by NSF
grant DMS-9205296
 and EC grant
CHRX-CT93-0411} \\
Helsinki University, Department of Mathematics,\\
Helsinki 00014, Finland\\ajkupiai@cc.helsinki.fi}

\date{}

\begin{document}

\maketitle \begin{abstract}

We explain how to apply  Renormalization Group ideas to the
analysis of the long-time asymptotics of solutions of partial differential
equations. We illustrate the method on several examples of nonlinear
parabolic equations. We discuss many applications, including the stability
of profiles and fronts in the Ginzburg-Landau equation, anomalous scaling
laws in reaction-diffusion equations, and the shape of a solution near a
blow-up point.

\end{abstract}

\section{Introduction.}

\vs{5mm}

The development of a qualitative theory of infinite dimensional dynamical
systems is a major scientific challenge. Such systems are expressed
through (nonlinear) partial differential equations, and we shall
concentrate
on equations of the form
\be u_t = \Delta u + F (u, \nabla u,
\nabla\nabla u). \en where, $u(x,t)\in {\bf R}$ (or ${\bf C}$ ), $x\in {\bf
R}^d$, $t\in{\bf R}^{+}$ and $F$ is nonlinear or random. Usually, the
following problems are considered: \begin{enumerate} \item[1.] Existence
and
regularity of the solution over a finite time interval. \item[2.] Extension
of the first problem to an infinite time interval. \item[3.] If the
solution
exists for all times, what is its asymptotic behaviour? How does the latter
depend on the initial conditions? If the solution ceases to exist after a
finite
time, we may still ask about its behaviour near the point where it breaks
down. \end{enumerate}

The first question is often easy to deal with. In particular, perturbation
methods (small initial data or small nonlinearity) tend to work well over
finite (or short) time intervals. However, naive perturbation theory  does
not, in general, yield an answer to the last two questions. A similar
situation was encountered in the theory of critical phenomena and in
quantum
field theory: perturbation theory works well whenever there are enough
cutoffs, but tends to diverge when the latter are removed. In order to
solve
these problems,  the Renormalization Group (RG) method was developed and
proved to be very useful. One of our goals is to develop RG ideas in the
theory of PDE's. Here, the role of the cutoff is played by the finite time.

The RG method for nonlinear PDE's was developed by Goldenfeld, Oono and
others  \cite{Go}. Previously, Barenblatt \cite{Ba} emphasized the role of
self-similar solutions of these equations. The latter can be viewed as
fixed
points of certain RG transformations (see below). Independently, we used RG
methods to deal with an equation like (1),with a linear but random $F$
\cite{BKRW2,BKRW1,BKRW3}. Here
we shall explain the basic RG ideas in the framework of PDE's, which turn
out
to be very natural and easy, and we shall review some of the rigorous
results
obtained so far.

The general idea behind the RG approach is to solve the problem
iteratively:
first integrate the equations far away from the cutoff to be removed, i.e.
here solve a finite time problem. Next, rescale the time and possibly the
space and the $u$ variables so as to produce a new problem similar to the
original one. Finally, iterate and see what happens. The method will work
when
the new problem tends to be simpler  than the original one
(meaning, e.g., a weaker nonlinearity). Upon iteration, it
will become simpler and simpler. Of course, this will happen only if we
have
chosen the scaling appropriately.

The underlying properties of the dynamical system that make the RG method
work are universality and scaling. In the PDE framework, scaling amounts to
the observation that many problems have solutions that behave
asymptotically
as \be u(x,t)\sim t^{-{\alpha\over 2}}f^* ({x\over{t^\beta}}) \label{1} \en
Often such a limit law is {\it universal}: the numbers $\alpha$, $\beta$
and
the function $f^*$ will not depend on the initial conditions or even on the
form of the equation. More precisely, pairs of initial data and equations
will fall into universality classes, corresponding to given $\alpha$,
$\beta$
and $f^*$. The fact that whole classes of problems may yield the same
asymptotic behaviour is called universality. In fact, $f^*$ will be a fixed
point of the RG transformation, while the exponents $\alpha$ and $\beta$
are
determined by the choice of the scaling. The basin of attraction of that
fixed point is the universality class. The RG will lead to a dynamical
system
picture, with an analysis of the stable, neutral and unstable manifolds of
the fixed points.

The outline of the paper is as follows: first, we give an elementary
definition of the RG transformation, starting from the (linear) heat
equation. Then, we show how the RG flow acts on spaces of initial data and
equations for various nonlinear perturbations of the heat equation, and for
data that decay at infinity. In Section 3, we consider similar questions,
but for data that do not decay at inifinity, which leads to the formation
of patterns and fronts. Finally (Section 4), we apply the RG method to
equations whose solutions blow-up in a finite time, and we give a list of
possible profiles near a blow-up point. We end this review with a list of
open problems.

\vs{5mm}

\section{The Renormalization Group.} \setcounter{equation}{0} \vs{5mm}

\subsection{The linear heat equation: Gaussian fixed points.}

To explain the RG idea, let us start with the simplest equation, the linear
heat equation: \be u_t = u_{xx} \en in one dimension, with integrable
initial data, $u(x,0) = f(x) \in L^1 ({\bf R})$. The solution is \be u(x,t)
= \frac{1}{\sqrt{4\pi t}} \int dy \exp (-\frac{(x-y)^2}{4t}) f(y) dy \en
{}From
this, it is easy to deduce that \be \sqrt t u(\xi \sqrt t, t) \to
\frac{e^{-\xi^2/4}}{\sqrt{4\pi}} \int f(y) dy
\en
as $t \to \infty$
(pointwise and in $L^1 ({\bf R}))$. Let us reformulate this (trivial)
result in
the RG language. Notationally, it is useful to take the initial time equal
to
1 instead of 0: \be u(x,1) = f(x) \en
This will not change anything to the
long-time asymptotics. Next, take a number $L>1$ and define the
renormalization map, acting on the initial data $f$ as \be (R_L f) (x) = L
u
(Lx,L^2) \en where $u(\cdot,L^2)$ is the solution of (1) at time $L^2$.
Thus,
we solve the equation up to a finite time $L^2$, rescale the solution and
take it as a new initial data. Here, $R_L$ is explicit: in Fourier
transform,
$e^{t \Delta}$ multiplies by $e^{-tk^2}$ and scaling $x$ by $L$ amounts to
scaling $k$ by $\frac{1}{L}$. Thus, \be
\widehat{R_Lf}(k)=e^{-k^2(1-L^{-2})}\hat{f}(L^{-1}k) \en (we solve (1) over
a
time interval of length $L^2-1$).

Observe that the set of multiples of \be \widehat{f^\ast_0} (k) = e^{-k^2}
\en forms a line of fixed points of the RG map (6). Note also that $R_L$
satisfies the semigroup property: Since $L$ is arbitrary, we may replace it
by $L^n$, and we have: \be R_{L^n} = R^n_L. \en This is simply because, if
we
define $u_L (x,t) = L u (Lx,L^2t)$, then \be (R_L f) (x) = u_L (x,1) \en
and
$u_L$ satisfies (1) again. Equation (1) is scale invariant.

Now, assume that our initial $f$ belongs to the basin of attraction of the
line of fixed points, i.e. \be R^n_L f \to Af^\ast_0 \en as $n\to \infty$,
for
some $A$, in a suitable space. The basic observation is that this is
equivalent to (3), by letting $t=L^{2n}$, $A=\hat{f}(0)$, and
combining (5,8,10). Note that
$A$ depends on the initial data and that this dependence is the only trace
left by $u(x,1)$ on the long-time asymptotics of the solution. This is why
the latter is called universal.

To see what kind of functions belong to the basin of attraction of this
line
of fixed points, write formally $\hat f (k) = \hat f (0) + ck + \cdots$.
Then, for $k$ of order one, we have from the scaling \be \widehat{R_Lf} (k)
=
\hat f (0) f^\ast_0 (k) + {\cal O} (L^{-1}) \en and, for large $k$, we get
contraction from the multiplication by $e^{-k^2(1-\frac{1}{L^2})}$ in (6).

To be more precise, we use below the following norm (inspired by
\cite{CEE}):
\be \| f \| = \sup_k (1+|k|^q) (|\hat f(k)| + |\hat f' (k)|) \en on a space
of functions with $\hat f \in C^1(R)$. We denote by ${\cal B}$ the
corresponding Banach space. The value of $q$ will be specified when we deal
with nonlinear problems. For the moment, any $q \geq 0 $ will do.
Any function $f\in {\cal B}$ can be written in Fourier transform as \be
\hat f
(k) = \hat f (0) e^{-k^2} + \hat g (k) \en with $\hat g (0) = 0$.
Then the convergence (10) with $A = \hat f (0)$ will follow from \\

\vs{3mm}

{\bf
Lemma 1} {\it There exists a constant $C$ independent of $L$ so that, if $g
\in
{\cal B}$ and $\hat g (0) = 0$,} \be \| R_Lg \| \leq C L^{-1} \| g \| \en
\vs{3mm}
{\bf Proof}\\ Write $\hat g (\frac{k}{L}) = \int^{k/L}_0 \hat g' (p) dp$
so,
$|\hat g (\frac{k}{L})| \leq \frac{|k|}{L} \| g \|$. The scaling (6) brings
a
factor of $L^{-1}$ in the derivative and $|k|^n e^{-k^2}$ is bounded by a
constant for any $n \geq 0$. \hfill$
\makebox[0mm]{\raisebox{0.5mm}[0mm][0mm]{\hspace*{5.6mm}$\sqcap$}}$ $
\sqcup$

\vs{2mm}

\noindent {\bf Remark 1.}  At this
point, we may comment on the choice of the constant $L$. It is arbitrary
except that we take it large enough so that $CL^{-1} < 1$ in (14). This
will
be assumed, wherever needed, about any other numerical constant entering
the
proofs. We shall denote generic constants by $C$, even when they change in
the same equation. The only important property of such constants is that
they
are independent of $L$.

\noindent {\bf Remark 2.} Using (13,14) we can prove (10), i.e. (3)
for $f \in {\cal B}$.

\subsection{Nonlinear heat equations.}

Let us now consider a less trivial example, namely
\be  u_t = u_{xx} + \lambda
u^p \en
where $\lambda $ could be scaled away but is introduced for
convenience. Throughout this Section, we work in one dimension (to simplify
the
notation), but the extension to higher dimensions is straightforward.
There is no problem
in proving the existence of solutions for any fixed time, provided
$\lambda$ is small
enough, and we may define the
renormalization group map as in (5). However, we do not have anymore the
semigroup property (8), because the equation is not scale invariant. In
fact,
$u_L (x,t) = L u (Lx,L^2 t)$ satisfies
\be
u_{Lt} = u_{Lxx} +\lambda
L^{3-p} u^p_L
\en
So, if we define $R_L$ on the pair $(f,\lambda)$ by \be R_L
(f,\lambda) = (u_L (x,1),\lambda L^{3-p}) \en we recover the semigroup
property (8), provided there is a set of pairs $(f,\lambda)$ that is mapped
into itself by $R_L$.

Now, consider $p>3$. Then the line of fixed points $(Af^\ast_0,0)$ of $R_L$
is stable. So we may expect the long time behaviour of the solutions of
(15)
to be again given by $Af^\ast_0$ for a suitable $A$. This is indeed the case.
However, the constant $A$ will not depend only on $f$, but also on $p$ and
$\lambda$.

We have the following result which is a variant of Theorem 1 in
\cite{BKL}. Fix $p \in {\bf N}, p>3$ in (15) and fix $q>1$ in (12); let
${\cal B}_1$ be the unit ball in the corresponding Banach space
${\cal B}$.

\vs{3mm}

{\bf Theorem 1} {\it There exist an $\epsilon >0$ such that, if $| \lambda
|
\leq \epsilon$ and $f \in {\cal B}_1$, equation (15) with $u(x,1) = f(x)$
has
a unique solution which satisfies, for some number $A = A (f,\lambda,p)$,
$$
\lim_{t \to \infty} \| \sqrt{t} u (\sqrt t\cdot, t) - A f^\ast_0 (\cdot) \|
=
0. $$ }

\vs{2mm}

\noindent {\bf Remark 1.}  For related results on the $p>3$ case,
 see \cite{CEE,Ga,GV,KP1}.

\noindent {\bf Remark 2.}  The convergence in ${\cal B}$ implies
convergence both in
$L^1$ and in $L^\infty$.

\vs{3mm}

Let us explain the main ideas of the proof (for details, see \cite{BKL}).
The
problem is to ``construct" the constant $A$ which is not given by an
explicit formula as in (3). This construction is done inductively.

First of all, the existence of the solution for finite times is rather
trivial (we take $\lambda$ small and $\| f \| \leq 1$). Indeed, we can
write
\be
u(t) = e^{(t-1)\partial^2} f + \lambda \int^{t-1}_0 ds e^{s \partial^2}
u^p (t-s) \en (leaving out the $x$ dependence) and use a standard fixed
point
argument (see \cite{BKL}; here we need only $q>1$, because there are no
derivatives in the nonlinear term). Since the first term in (18) is the
linear evolution, we may write \be R f = R_0 f + v \en where $R_0$ denotes
the linear RG map (5) (we suppress here the $L$ dependence of $R$) and $\|
v
\|$ is ${\cal O} (\lambda)$. Now, write \be f = A_0 f^\ast_0 + g_0 \en
with
$\hat g_0(0) =0$. This splitting of $f$ is not preserved by $R$, as it was
in
the linear case. Thus, we shall have to change $A$ at each step. Write \be
R
f = A_1 f^\ast_0 + g_1 \en
where again $\hat{g_1} (0) = 0$. From (19) and
(20), we see that
\be A_1 = A_0 + \hat v (0), \en
since $\widehat{R_0 g_0}
(0) = 0$ (see (6)), and
\be g_1 = R_0 g_0 + v - \hat v (0) f^\ast_0. \en
Using Lemma 1
and $\| v \| = {\cal O}(\lambda)$, we have
\be
A_1 = A_0 + {\cal O}(\lambda)
\en
and
\be \| g_1 \| &\leq & CL^{-1} \| g_0 \| + {\cal O}(\lambda) \nonumber \\
&\leq & L^{-(1-\delta)} \en for any $\delta > 0$, provided $\lambda$ is
small
enough, since $\| g_0 \| = {\cal O}(1)$ for $f \in {\cal B}_1$.

Now iterate: under scaling $\lambda$ becomes $\lambda_1 = \lambda L^{3-p}$,
and after $n$ steps \be \lambda_n = \lambda L^{(3-p)n}
\en
Thus, $\| v_n \|$
will be ${\cal O} (\lambda_n)$ for the $n^{th}$ iteration, and, writing
$R^n f = A_n f_0^\ast + g_n$, we have
\be
A_{n+1}
&=& A_n + \hat v_n(0) \\ g_{n+1} &=& R^0_L g_n + v_n - \hat v_n (0)
f^\ast_0,
\en
with the bounds: \be |A_{n+1} - A_n| \leq {\cal O} ( \lambda_n) \en
\be \| g_{n+1} \| &\leq & CL^{-1} \| g_n\| + {\cal O} (\lambda_n)
\nonumber\\
&\leq & L^{-(1-\delta)n} \en (use (26) and $p-3 \geq 1$ since $p$ is an
integer).

Therefore there exists a constant $A$ such that $A_n \to A$ as $n \to
\infty$
and \be |A - A_n| \leq {\cal O} (\lambda_n). \en
Then, \be R^n f &=& A_n
f^\ast_0 + g_n \nonumber\\ &= & A f^\ast_0 + (A_n - A) f^\ast_0 + g_n \en
and
\be \| R^n_L f - A f^\ast_0 \| \leq L^{-(1-\delta)n} + {\cal O} (\lambda_n)
\en which goes to zero as $n \to \infty$. Using (5,8,17),
this proves the claim, at least for a special sequence of times;
the extension to all times is easy \cite{BKL}.

\vs{2mm}

\par\noindent {\bf Remark 1}.  From (33,26),
one gets an estimate ${\cal O} (t^{-(\frac{1-\delta}{2})})$ on the rate of
convergence.
\par\noindent {\bf Remark 2.} The sign of $\lambda$ is not important here.
It
matters only for $p \leq 3$, see below.
\par\noindent {\bf Remark 3.} $R_0$ is the linearization
of the component of the operator $R$ that acts on initial data, around the
fixed point $f^\ast_0$. All functions of the form $k^n e^{-k^2}, n \in N$,
are (Hermite) eigenvectors: \be R_0 k^n e^{-k^2} = L^{-n} k^n e^{-k^2}
\en
The fact that the scaling transforms the linear
evolution semigroup, $e^{t \partial^2}$, which has continuous spectrum,
into
an operator with discrete spectrum, is one of the interesting features of
this
method. Using it, one can extend the above analysis and obtain higher order
asymptotics of the solution in inverse powers of $t$, see
subsection 2.5
below, and \cite{Li,Wa}.

\subsection{The RG flow in spaces of equations and data.}

Of course, Theorem 1 can be extended to a much more general class of
nonlinearities. Consider an equation of the form \be  u_t = u_{xx} + F
(u,u_x,u_{xx}) \en Then, the map $R_L$ will be defined in general by \be
R_L =
(L^\alpha u_L (\cdot,1),F_L) \en where \be F_L (a,b,c) = L^{2+\alpha} F
(L^{-\alpha} a, L^{-1-\alpha} b, L^{-2-\alpha} c) \en If we consider a
monomial $F(a,b,c) = a^{n_1} b^{n_2} c^{n_3}$, and the previous scaling
$\alpha = 1$, we have $F_L = L^{-d_F} F$ with \be d_F = n_1 + 2 n_2 + 3 n_3
-
3 \en and using standard RG terminology, we call the monomial  {\it
relevant} if $d_F < 0$, {\it marginal} if $d_F = 0$, and {\it
irrelevant} if $d_F > 0$.

Considering $n_i \geq 1$ or $n_i = 0$, the nonlinearities are easily
classified:
$u^p$ is relevant for $p<3$; $u^3$ and $u u_x$ are marginal, and
everything else is irrelevant.
Then, Theorem 1 extends to the irrelevant case, provided we take
$q>3$ in (12) and $F: {\bf C}^3
\to {\bf C}$ analytic in a neighbourhood of zero \cite{BKL}.
\vs{2mm}
\par\noindent {\bf Remark.} More general irrelevant nonlinearities
(nonlocal,
integro-differential terms) were considered by Taskinen \cite{Ta}. Besides,
as mentioned
in \cite{BKL}, we may replace the second derivative in (15) by
$-(-\Delta)^{\beta/2}$ and obtain
similar results.

A nice extension to the study of waves satisfying a generalized KdV-Burgers
equation
was made in \cite{Wa1}. There, the highest order (third) derivative turns
out to be
irrelevant.

\vs{3mm}

The two marginal terms exhibit different behaviours. For $u^3$, it is
essential to replace the plus sign by a minus one (with $\lambda >0$) in
equation (15) (with a plus sign, and $p\leq 3$, any positive initial data
leads to
a solution that belows up in a finite time \cite{Fu,Le}). The nonlinearity
turns out to be irrelevant in the next
order. The coefficient $A_n$ satisfies a recursion of the form $A_{n+1} =
A_n
- \lambda \beta A^3_n$ + higher order terms, where $\beta > 0$ is an
explicit
constant. This implies that $A_n \sim n^{-\frac{1}{2}}$ (here we use
$\lambda$
positive) and since $t = L^{2n}$, this translates into the following
logarithmic correction (see \cite{BKL}): \be u (\xi \sqrt t , t) \simeq (t
\log t)^{-\frac{1}{2}} f^\ast_0 (\xi) \en

On the other hand, for $u u_x$, we have Burgers' equation  and, via the
Cole-Hopf transformation, we get a new line of fixed points: \be f^\ast (x)
=
\frac{d}{dx} \log (1+ A e(x)) \en
where $e(x) = \int^x_{-\infty} e^{-
\frac{y^2}{4}} dy$ is the error function and $A$ is a parameter. A theorem
analogous to Theorem 1 can be proven for these new fixed points
\cite{BKL}. Of course we can also add to these marginal terms a general
irrelevant $F$ and obtain a similar result.

\subsection{Non-Gaussian fixed points.}

Let us now turn to the relevant case, $u^p, 1 < p < 3$. Actually, since we
will not restrict ourselves to integer $p$ or to everywhere positive $u$'s,
we
shall consider the equation:
\be u_t = u_{xx} - u |u|^{p-1} , \en
but the
minus sign will  again be essential. Here, the parameter
$\lambda$ of (15) has been eliminated by a rescaling of $u$.

Since the nonlinear term is relevant, the Gaussian fixed point
 will be of no use. In order to find another fixed point, consider the
transformation (36) with $\alpha= {2\over p-1}$. The reason for this choice
is that equation (41) is then scale invariant: $F_L=F$. It is easy to see
that we shall have a fixed point of the transformation if we have a
solution
of the form: \be u(x,t) = t^{- \frac{1}{p-1}} f (xt^{- \frac{1}{2}}) \en
where $f(\xi)$ is a function of one variable which solves: \be f'' +\ha \xi
f' + {f\over p-1} -f^p = 0. \en
These solutions are called self-similar. The
theory of positive solutions of (43) has been developed in
\cite{Bre,Ga,KP1}.
The main result is that, for any $p>1$, there exist smooth, everywhere
positive solutions, $f_\gamma$, of (43) with $f_\gamma^{'}(0)=0$ and
$f_\gamma(0) =\gamma$ for $\gamma$ larger than a certain critical value
$\gamma_p$ (but not too large).  The decay at infinity of these solutions
is
given by
\be f_{\gamma}(x) \sim |x|^{- \frac{2}{p-1}} \en
as
$|x|\rightarrow\infty$ if $\gamma > \gamma_p$, while, for $\gamma=
\gamma_p$,
it decays at infinity as \be f_{\gamma_p}(x) \sim |x|^{\frac{2}{p-1}-1}
e^{-
\frac{x^2}{4}}. \en
\vs{2mm}
\par\noindent {\bf Remark.} The existence of a critical $\gamma_p$ can be
understood intuitively by viewing (43) as Newton's equation for a particle
of
mass one, whose ``position" as a function of ``time" is $f(\xi)$. The
potential is then $U(f) = \frac{f^{2}}{2(p-1)} - \frac{f^{p+1}}{p+1}$ and
the
``friction term" $ \frac{\xi}{2} f^{'}$ depends on the ``time" $\xi$.
Hence,
if $f_\gamma^{'} (0) = 0$ and $f_\gamma(0)=\gamma$ is large enough, the
time
it takes to approach zero is long and, by then, the friction term has
become
sufficiently strong to prevent ``overshooting". However, as $p$ increases,
the potential becomes flatter and one therefore expects $\gamma_p$ to
decrease with $p$.

\vs{3mm}

We shall now explain why these self-similar solutions are stable. Consider
the
initial data (taken as before at time 1 for convenience)
\be
u(x,1)=f_\gamma(x)+h(x) \en
with $\gamma\geq\gamma_p$ and $h\in B$, where $B$
is the Banach space of $L^\infty$ functions equiped with the norm (with
some
abuse of notation!)
\be \| h \|_\infty = {\rm ess}\sup_\xi |h(\xi)
(1+|\xi|^q)|. \en
One can show the following \cite{BK5}:

\vs{3mm}

\no {\bf Theorem 2} {\it Let $1 < p < 3 $. There exist $\varepsilon > 0, C
<
\infty$ and $\mu > 0$ such that, if the initial data $u(x,1)$ of} (41)
{\it is given by} (46) {\it with $h\in B$ and satisfies

$$ \|h\|_\infty \leq \varepsilon, \; or \; h \geq 0 (a.e.) $$
then,} (41) {\it has a unique classical solution and, for all $t$, $$ \|
t^{\frac{1}{p-1}} u(\cdot t^{\frac{1}{2}},t) - f_\gamma(\cdot) \|_\infty
\leq
C t^{- \mu}\| h\|_\infty $$}

\vs{2mm} \no {\bf Remark.} For related results on the stability of
self-similar solutions, see \cite{EK1,EK2,EK3,Ga,KP2}.

\vs{3mm}

Let us explain the main ideas of the proof. Given the initial data (46), it
is convenient to rewrite (41) in terms of the variables $\xi = xt^{-
\frac{1}{2}}$ and $\tau = \log t$; so, define $v(\xi, \tau)$ by: \be u(x,t)
=
t^{- \frac{1}{p-1}} (f_\gamma (xt^{- \frac{1}{2}}) + v (xt^{- \frac{1}{2}},
\log t)) \en where now \be v(\xi,0) = h(\xi). \en
Then, (41) is equivalent to
the equation
\be
v_\tau  = {\cal L} v -
\left(|f_\gamma+v|^{p-1}(f_\gamma +v) - f_\gamma^p - pf_\gamma^{p-1} v
\right) \equiv {\cal L} v + N(v) \label{123} \en where we used the fact
that
$f_\gamma$ solves (43) and we gathered the linear terms in
$$ {\cal L} = {\cal
L}_0 + V_\gamma, $$
with
\be {\cal L}_0= \frac{d^2}{d\xi^2} + \frac{\xi}{2} \frac{d}{d\xi} +
\frac{1}{p-1}, \label{11} \en
and \be V_\gamma(\xi)=-pf_\gamma^{p-1}(\xi).
\label{124} \en

To prove the theorem, it suffices to show that the corresponding solution
of
(50) goes to zero as $\tau \rightarrow \infty$. For that, the main estimate
will be that the semigroup $e^{\tau{\cal L}}$ contracts, at least for
$\tau$
large: \be \|e^{\tau{\cal L}} \|\leq Ce^{-\mu\tau} \en for some $\mu>0$,
$C<\infty$. Given (53), the control over the nonlinear term in (50) is
standard.

There are two important ingredients in the proof of (53). The first is the
fact that $e^{\tau{\cal L}}$ is a contraction in a suitable Hilbert space
of
rapidly decreasing functions. To see this, note first that ${\cal L}_0$ can
be
conjugated to the following Schr\"odinger operator:
\be
e^{\frac{\xi^2}{8}} {\cal L}_0
e^{- \frac{\xi^2}{8}} = \frac{d^2}{d\xi^2} - \frac{\xi^2}{16} - \frac{1}{4}
+
\frac{1}{p-1},
\en
i.e. to minus the Hamiltonian of the harmonic oscillator. Thus, ${\cal L}_0$
is self-adjoint on its domain ${\cal D}({\cal L}_0) \subset L^2 ({\bf R},
d\mu)$, where $$ d\mu(\xi)=e^{\frac{\xi^2}{4}}d\xi. $$ ${\cal L}_0$ has a
pure point spectrum $\{{1\over p-1}-{1\over 2} -{m\over 2}\;|\;
m=0,1,\dots\}$ and the largest eigenvalue $ {1\over p-1}-{1\over 2}$ is
{\it
positive} if $1<p<3$. Thus $e^{\tau{\cal L}_0}$ is {\it not} contractive
and,
for $e^{\tau({\cal L}_0+V_\gamma)}$ to contract, we need to use the
potential
in a non-trivial way (this is the reason why $1<p<3$ is harder than the
$p>3$
case).

Remarkably, it is possible to prove that ${\cal L}< -E <0$ without a
detailed
study of the function $f_\gamma$, but only using equation (43). Indeed, one
first shows that $\cal L$ is self-adjoint and that $-E_\gamma$, its largest
eigenvalue, satisfies $-E_\gamma\leq -E_{\gamma_p}$. So, writing $E \equiv
E_{\gamma_p}$, it is enough to show that $-E <0$. Next, it is easy to see,
using the Feynman-Kac formula \cite{Si} and the the Perron-Frobenius
theorem
\cite{GJ} that ${\cal L}$ has a unique eigenvector $\Omega$ with eigenvalue
$-E$ and $\Omega$ can be chosen to be strictly positive. Write (43) as $$
{\cal L} f = -(p-1) f^p. $$ So, $$ (\Omega, {\cal L} f) = -(p-1)
(\Omega,f^p)
$$ where $(\cdot,\cdot)$ denotes the scalar product in $L^2 ({\Bbb
R},d\mu)$.
By the self-adjointness of ${\cal L}$ and the definition of $\Omega$, i.e.
${\cal L}\Omega = -E\Omega$, we have $$ -E = -(p-1)
\frac{(\Omega,f^p)}{(\Omega,f)} < 0 $$ since $\Omega$ and $f$ are strictly
positive. This proves our claim.

Notice that functions in $L^2 ({\bf R}, d\mu)$ have essentially a Gaussian
decay at infinity, which is much faster than what is allowed in our Banach
space $B$, see (47). An extra work is needed, using the Mehler's formula
for
the kernel of $e^{\tau{\cal L}_0}$ (see \cite{BK5} for details).

Finally, let us express this result in the RG language. The RG map has a
fixed point ${ R}_{L}(f_\gamma,F^*)=(f_\gamma,F^*)$ for
$F^*(u)=-u|u|^{p-1}$.
There are different fixed points for different values of $\gamma$, but
since
the decay of the functions in $B$ is faster than the one of any of these
fixed points, each of them is unique in the corresponding set $\{f_\gamma +
h\}$ for $h$ as in Theorem 2.
In other words, unlike the situation for $p>3$, we do not have to deal with
a
line of fixed points and here there is no constant like $A$ in Theorem 1.
This
is reflected in the fact, (53), that the linear semigroup $e^{\tau{\cal
L}}$
contracts.

\vs{2mm} \no {\bf Remark.}
The reader who is familiar with the theory of critical phenomena will
recognize an analogy between the behaviour of an Ising model at its
critical
point, when the dimension of the lattice changes and the asymptotic
behaviour
of the solution of (41) when $p$ varies. This is further discussed in \cite
{BKL}. Here, our results for $p<3$ are non-perturbative, i.e. we do not use
any ``$\varepsilon$-expansion" (see \cite{Wa} for an analysis of the
bifurcation
at $p=3$).

Also, the reader may notice that our approach to the renormalization group
is close
to the ``Wilsonian" one in the theory of critical phenomena, while the
method developed
in \cite{Go} is analogous to the renormalized perturbation of quantum field
theory.
For a rigorous treatment of Barenblatt's equation, which was one of the
first example
analyzed in \cite{Go}, see \cite{KP3}.

\subsection{Higher-order asymptotics.}

It is interesting to reexamine the $p\geq 3$ case, using what  we learned
from the analysis of $p<3$. This has been done by Wayne \cite{Wa}.
Namely, consider equation (15), and make the  change of variable (48), but
replacing $f_\gamma$ by 0,
i.e. \BE u(x,t) = t^{-\frac{1}{p-1}} v(xt^{-1/2},\log t) \EN Then
$v(\xi,\tau)$ satisfies
\BE
v_\tau  = {\cal L}_0 v + \lambda v^p \EN with
${\cal L}_0$ as in (51); we have computed the spectrum of  ${\cal L}_0$
which
is  $ \{ \frac{1}{p-1} - \frac{1}{2} - \frac{m}{2} | m=0,1,\cdots\}. $ Now,
notice that, for $p>3$, this spectrum is entirely negative.  Thus
$e^{\tau{\cal L}_0}$ contracts exponentially and it is  not hard
to show
that, for $\lambda$  small, and $v(\xi,0)$ bounded
 in a suitable norm,  the same contraction holds
for the solution of (56). Actually,  since the largest eigenvalue of ${\cal
L}_0$ corresponds  to $m=0$, we expect a decay like
$e^{\tau(\frac{1}{p-1}-\frac{1}{2})}$, which,
 using $\tau = \log t$ and combining it with the $t^{-\frac{1}{p-1}}$
prefactor in (55), leads to a $t^{-1/2}$ decay of $u(x,t)$, for all  $p\geq
3$. This of course has to be the case, in view of Theorem 1.

Actually, much more can be said. Take the $n$ largest eigenvalues
$\{\lambda_1,\cdots,\lambda_n\}$ of ${\cal L}_0$ and the corresponding
subspaces. Then, using a theorem of Gallay \cite{Gal2}, Wayne constructs
an
invariant manifold (in the $v$ variable) such that any solution  (for $p>3$
and small initial data) will converge to that invariant  manifold at a rate
at least of order  $t^{-(\lambda_{n+1}-\delta)}$ (with, as before, $\delta$
arbitrarily  small for sufficiently small initial conditions).

Since, for $n=1$, the eigenvector of ${\cal L}_0$ is $e^{-\xi^2/4}$,  one
easily recovers Theorem 1. But all higher order asymptotics  in time of
$v(\xi,\tau)$, hence of $u(x,t)$, can also be obtained  in terms of Hermite
functions (such higher order corrections were  also derived in \cite{Li}).

For $p<3$, an invariant manifold can still be constructed but it  becomes
unstable $(\frac{1}{p-1} - \frac{1}{2} >0)$ and thus, it does not, in
general, give
information on the long-time asymptotics.  Of course, we know from Theorem
2
what the situation is: instead  of 0, we have to consider the solution
$f_\gamma$ and,  presumably, a similar picture (with invariant manifolds
and
higher  order asymptotics) holds there, since the spectrum of the operators

 ${\cal L}$ for $p<3$ is qualitatively similar to the one of ${\cal L}_0$
for $p>3$.

\subsection{An application to reaction-diffusion equations.}

In \cite{BX}, Berlyand and Xin consider the following model which  leads to
a
nice RG analysis and which exhibits the novel feature  of anomalous
exponents
depending on the initial data: \BE u_t &=& \Delta u + v u^{p-1}  \\ v_t &=&
\Lambda^{-1} \Delta v - v u^{p-1} \EN where $u=u(x,t), v = v (x,t), x \in
{\bf R}^n$ (we take $n=1$ for simplicity)  and $\Lambda > 0$ is the Lewis
number. Equations (57,58) model  a chemical reaction $A \to B$ where $v$ is
the
mass fraction  of reactant $A$ and $u$ the one of reactant $B$.

First, observe that for $p>3$, the analysis done for (15) applies  and the
asymptotics of $u,v$ is Gaussian. So let $p=3$, so that  the nonlinear
terms
are marginal, and let us see heuristically  what happens. If $u$ and $v$
are
positive (which is physically  necessary and is preserved by (57,58) due to
the
maximum principle)  then, again by the maximum principle, $u$ is larger
than
the solution  of the heat equation with the same initial data: \BE u(x,t)
\geq \frac{A}{t^{1/2}} e^{-\frac{x^2}{4t}}. \EN where $A$ depends on
$u(x,0)$.

Then, $v$ is less than the solution of (58) with $u^{p-1} = u^2$
replaced by its lower bound (59). Calling $\bar v$ this upper  bound on
$v$,
$\bar v$ solves a linear equation
\BE \bar v_t = \Lambda^{-1} \Delta \bar v -
A^2 t^{-1} e^{-\frac{x^2}{2t}} \bar v.
\EN
It is not hard to see,  by direct substitution, that (60) admits a
self-similar solution of the form
\BE \bar v (x,t) = t^{-\frac{\alpha}{2}}
f^*_\alpha (\frac{x}{\sqrt t}). \EN
Indeed, writing $f^*_\alpha (\xi) =
e^{-\frac{\Lambda \xi^2}{8}} \Psi (\xi)$,  $\Psi$ solves \BE - \Lambda^{-1}
\Psi'' + (\frac{\Lambda}{16} \xi^2 +  \frac{A^2}{4 \pi} e^{- \xi^2/2} +
\frac{1}{4}) \Psi = \frac{\alpha}{2} \Psi \EN We see that, requiring
$f^*_\alpha$ to be positive means that $\Psi$ is the  ground state of the
operator in the LHS of (62) i.e. of an harmonic  oscillator perturbed by a
positive, rapidly decaying, potential.  The exponent $\frac{\alpha}{2}$ is
the corresponding ground  state energy. For the harmonic oscillator,
$\frac{\alpha}{2} = \frac{1}{2}$  and, perturbation theory tells us that,
for $A$ small,
\BE
\frac{\alpha}{2} = \frac{1}{2} + \frac{A^2}{4 \pi \sqrt{2\Lambda^{-1}+1}}  +
h.o.t. \; \mbox{in} \; A \; > \; \frac{1}{2}. \EN
Thus,
$\alpha$ depends on $A$, i.e. on $u(x,0)$. Since $\alpha > 1$,
 $\bar v$, and hence $v$, decay strictly faster than the  solution of the
heat equation. Inserting this in (57) gives an upper solution
for $u$, which solves
an equation where the
nonlinear term is now {\it irrelevant}.

These  considerations lead us to
expect $u$ to have the heat  equation decay: \BE u(x,t) \simeq
\frac{A}{\sqrt
t} e^{-x^2/4t} \EN and $v$ to have the anomalous decay:
\BE v(x,t) =
\frac{B}{t^{\alpha/2}} f^*_\alpha (\frac{x}{\sqrt t}) \EN
where $A,B$ and
$\alpha$ depend on the initial data. Upper and  lower bounds of the form
(64,65) are proven in \cite{BX}. From the RG point  of view, one can define
a
map $$ R_{L,\alpha} (u,v) = (u_L(\cdot,1),v_L(\cdot,1)) $$ where \BE
u_L(x,t) &= & Lu(Lx,L^2t), \\ v_L (x,t) &=& L^\alpha v(Lx,L^2t). \EN We
have thus
a two-parameter family of fixed points $(Af^*_0,B f^*_{\alpha(A)})$, and
the scaling exponent varies continuously with $A$.

\section{Patterns and fronts.}
\setcounter{equation}{0}
 \vs{5mm}
\subsection{The Ginzburg-Landau equation.}

In the previous section, we have seen that nonlinear equations  with
initial
data decaying at infinity produce universal,
Gaussian or non-Gaussian,
diffusive profiles. Here we shall show that other types of asymptotic
behaviour can also exhibit such universality. To discuss a concrete
example,
consider  the Ginzburg-Landau equation \be u_t= u_{xx}+u-|u|^2u \en where
$u:{\bf R}\times{\bf R}\rightarrow{\bf C}$ is complex. This equation has a
two parameter family of stationary solutions  \be
u_{q\theta}(x)=\sqrt{1-q^2}e^{i(qx+\theta)} \en and a natural question is
to
inquire about the time development of initial data $u(x)$ which approach
two
such solutions at $\pm\infty$: \be
\lim_{x\rightarrow\pm\infty}|u(x)-u_{q_\pm\theta_\pm}(x)| =0 . \en

This problem has been extensively studied for equation (1) with $u$ {\it
real} and $u(-\infty)=1$, $u(\infty)=0$ (i.e. $q_-=0$, $q_+=0$)
\cite{AW,Br}.
Then the solution takes the form of a propagating front. This
occurs because
$u(x)=1$ is a stable stationary solution, while $u(x)=0$ is an unstable
one.
For complex $u$, the solutions (2) are stable, under small perturbations,
for
$q^2< {1\over 3}$ (the Eckhaus stable domain) \cite {CEE}.

We shall consider
two questions. The first one was suggested in \cite{CE}, namely we take
$q_\pm$ in (3) small, belonging to the Eckhaus stable domain, but not
necessarily equal. What is then the long-time  asymptotics of the solution?
The second question concerns the stability of the real front solutions  of
(1) for real or complex $u$. For reviews on these questions, we refer the
reader to
\cite{CE1,CE2}.

\subsection{Non-Gaussian patterns.}

For the first question, we considered in \cite{BK2} a class of initial data
satisfying (3) and we showed that, for any interval $I$, \be \sup_{x\in
I}|u(x,t)- e^{i\sqrt{t}\phi^*}u_{q^*\theta^*}(x)|\leq {C_I\over\sqrt{t}}
\en
where the constants $q^*$, $\phi^*$ and $\theta^*$ depend only on the
boundary conditions (3). For a more detailed bound, see Section 3 of
\cite{BK2}. Graphics of the solution can be found in \cite{CE2}.

In order to see where (4) comes from, we write, following \cite{CE}, \be
u=(1-s)e^{i\phi} . \en Equation (1) becomes, in these variables,
\be s_t&=&
s_{xx}-2s+3s^2-s^3+ \phi_x^2-s\phi_x^2\equiv s_{xx}-2s+F(s,\phi_x)\non\\
\phi_t&=& \phi_{xx}-{2\phi_x s_x\over 1-s}\equiv \phi_{xx}+G(s,
s_x,\phi_x) \en with the initial data ( taking again $t=1$ as initial time)
\be \lim_{x\rightarrow\pm\infty}s(x,1)=s_{\pm}\; ,\;\;
\lim_{x\rightarrow\pm\infty}|\phi(x,1)-\phi_\pm x-\theta_\pm| =0 . \en
where
$2s_\pm=F(s_\pm,\phi_\pm)$. We proved in \cite{BK2} the following
asymptotics,
as $t\rightarrow\infty$:
 \be &&\phi(x,t)=\sqrt{t}\phi^*({x\over\sqrt{t}})+
\theta^*({x\over\sqrt{t}})+{\cal O}({1\over\sqrt{t}}) \\&&s(x,t)=
s^*({x\over\sqrt{t}})+{1\over\sqrt{t}} r^*({x\over\sqrt{t}})+{\cal
O}({1\over
t}) \en for a set of initial data satisfying (7), with $\phi_\pm $, and
$\theta_\pm$ small enough (see again \cite{BK2} for the precise definition
of
the norms in which these limits take place).  The functions $\phi^*$,
$\theta^*$, $s^*$, $r^*$ are {\it universal}, depending on the initial data
only through the boundary conditions (7). They are smooth and therefore
(expanding $\phi^*$ to first order, and using (2)) the phase of $u$ will
have
the asymptotics (4), with
$$\phi^*=\phi^*(0),\;\;q^*=\phi^{*'}(0),\;\;\theta^*=\theta^*(0).$$
The behaviour of $s$ will be discussed below.

The peculiar scaling behaviour exhibited by (8) can already be understood
through the linear heat equation. Previously, we considered the Gaussian
fixed point of the RG in a space of integrable initial data. Since the heat
equation has stationary solutions of the form $a+bx$, we may also consider
the solution
of $\phi_t = \phi_{xx}$ with initial data $\phi (x,1) = f(x)$ satisfying
(7). The
solution can be computed:
\be
t^{-\ha}\phi(\sqrt{t}x,t+1)={1\over \sqrt{4\pi t}}\int e^{-{1\over 4}
(y-x)^2} f(\sqrt{t}y)dy,\non\\ \rightarrow_{t\rightarrow\infty}
\phi_{-}x+(\phi_{+}-\phi_{-})(x+2{d \over dx})e(x)
\equiv \phi^*_0(x) \en
where $e(x) = \int^x_{-\infty} e^{-
\frac{y^2}{4}} dy$ is the error function.
This gives the first term in (8), with $\phi^\ast$ replaced by
$\phi^\ast_0$. If
we want to obtain the next correction, we write $\phi^\ast =
\phi^\ast_0 + \theta$,
where $\theta (x,t)$ solves $\theta_t = \theta_{xx}$ with $\theta(x,1) \to
\theta_{\pm}$ as
$x \to \pm \infty$. One computes that
\be
\theta(x,t) \to \theta^\ast_0 \left(\frac{x}{\sqrt t}\right) = \theta_- +
(\theta_+ - \theta_-) e \left(\frac{x}{\sqrt t}\right)
\en
and this yields the second term in (8). It is trivial to check that
$\phi^\ast_0$ (resp. $\theta^\ast_0$) is
a fixed point of the RG map (2.36) with $\alpha = - 1$ (resp. $\alpha =
0$), and $F=0$.

We shall explain below that $\phi^\ast, \theta^\ast$ are small
perturbations of $\phi^\ast_0, \theta^\ast_0$.
However, as we saw, the scaling (8,4) can be understood qualitatively on
the basis of the heat
equation, and the boundary conditions (7). We still need to understand (9).

The latter comes from the ``slaving" of $s$
to $\phi$, due to the linear $-2s$ term in (6). This
will imply that $s$ will have the right form so that (4) holds.
Concretely,
the slaving means that, to leading order, we may solve the algebraic
equation $-2s+F(s,\phi_x)=0$, which is equivalent to $1-s = \sqrt{1-(
\phi_x)^2}$, and substitute the result in the equation (6) for $\phi$.
Since the derivative of
$s$ is proportional to the second derivative of $\phi$, we obtain  an
equation for $\phi$ only, of the form: \be
\phi_t=(1+a(\phi_x,\phi_{xx}))\phi_{xx}\; ,\;\;\phi(x,1)=f(x) \en with the
boundary conditions (7) for the initial data $f$. The function $a$ is
analytic around the origin.

The RG map acts again on pairs of  initial data and
equations and can be defined (with $\alpha=-1$) as \be R_L(\phi,a) =
(\phi_L,
a_L) \en where
\BE \phi_L(x,t)=L^{-1} \phi(Lx,L^2t) \nonumber
\EN and \be
a_L(u,v)=a(u,L^{-1}v). \en
The semigroup property follows by observing that $\phi_L$ satisfies
\be
\phi_{Lt}=(1+a_L(\phi_{Lx},\phi_{Lxx}))\phi_{Lxx}. \en

We may now iterate $R$ as before, i.e. solve a finite time problem, to
study
the asymptotics of (12). One wants to show that
there exist functions
$\phi^{\ast}, a^{\ast}$ such that \be R_L^n(\phi, a)
\rightarrow (\phi^\ast, a^{\ast}) \en
 where \be
R_L(\phi^\ast, a^{\ast}) =(\phi^\ast, a^{\ast})
 \en is the fixed point of the RG,
corresponding to the scale-invariant equation
$ \phi_t=(1+a^\ast) \phi_{xx}$.  Then, replacing
${x\over\sqrt{t}}$ by $\xi$, the asymptotics of the
original problem is given by
\be \phi(\xi t^{1/2},t)\sim t^{{1\over
2}}\phi^\ast (\xi). \en

Because of the factor $L^{-1}$ in the second argument of $a$ in (14), the
fixed point equation is
\be \phi_t=(1+a^*(\phi_x))\phi_{xx} \en where
$a^*(\cdot) =a(\cdot,0)$. The fixed point is the scale invariant solution $
\phi(x,t)=\sqrt{t}\phi^*({x\over\sqrt{t}}). $ We get for $\phi^\ast$ the
equation
 \be (1+a^*){d^2 \over d\xi^2}\phi^* +\ha \xi{d
\over d\xi} \phi^*-\ha\phi^*=0 \en with $a^*=a^*({d\over d\xi} \phi^*)$
and,
for small $\phi_\pm$, we look for a solution \be \phi^*=\phi^*_0+\rho \en
where $\rho (\pm\infty)=0$ and $\phi^*_0$ is the ``Gaussian" solution (10),
which solves (20) with $a^*=0$. This is easy to solve
by a fixed point argument(see Proposition 1 in
\cite{BK2} or the Proposition in Section 4 of \cite{BKL}).

This gives us the first term in (8). Turning to $\theta^*$, we write
\be \phi(x,t)=\phi^*(x,t)+\theta(x,t) \en where, with an abuse of
notation, $\phi^* (x,t)= \sqrt{t} \phi^* (\frac{x}{\sqrt{t}})$ and $\phi^*$
is given by (21), while $\phi$ solves (12). Then,
\be
\theta_t=\theta_{xx}+(a\phi_{xx}-a^*\phi^*_{xx}) \en with
$\theta(\pm\infty)=\theta_\pm$. Now we set
\be \theta_L(x',t')=\theta(Lx',L^2t').
\en
Putting $x=Lx', t=L^2t'$, we have  $\frac{d^l}{dx^l} \phi^*
(Lx',L^2t')=L^{1-l}
\frac{d^l}{dx'^l} \phi^*(x',t')$ and therefore $\theta_L$, satisfies
the equation (replacing $(x',t')$ by $(x,t)$):
\be
\theta_{Lt}&=&\theta_{Lxx}+L(a(\phi^*_x+L^{-1}
\theta_{Lx}, L^{-1}\phi^*_{xx}+L^{-2}\theta_{Lxx})
(\phi^*_{xx}+L^{-1}\theta_{Lxx}) \non\\
&&-a^*(\phi^*_x)\phi^*_{xx}). \en
Thus,
reasoning as above, we expect \be \theta_{L^n}\rightarrow\theta^* \en where
$\theta^*(x,t)=\theta^*({x\over \sqrt{t}})$ satisfies the
$L\rightarrow\infty$ form of (25): \be
\theta^*_t=\theta^*_{xx}+a^*\theta^*_{xx}+
(a_u(\phi^*_x,0)\theta^*_x+a_v(\phi^*_x,0)\phi^*_{xx})\phi^\ast_{xx}.
\en
This is a linear equation, easy to solve, whose solution is, for
$\theta_\pm$
small, a small perturbation of the ``Gaussian" solution (11) (which solves
(27)
with $a=0$).
Finally, one sets (with the same abuse of notation)
\be
\phi(x,t)=\phi^*(x,t)+\theta^*(x,t)+\psi(x,t)=\sqrt{t}\phi^*
({x\over\sqrt{t}})+\theta^* ({x\over\sqrt{t}})+\psi(x,t) \en

As for the $s$
variable, one gets \be s(x,t)=s^*(x,t)+r^*(x,t)+v(x,t) \en where \be
s^*(x,t)=s^*({x\over\sqrt{t}}),\;\;
r^*(x,t)={1\over\sqrt{t}}r^*({x\over\sqrt{t}}) \en are fixed points
``slaved'' respectively to $\phi^*$ and $\theta^*$. They satisfy the
boundary
conditions \be &&\lim_{x\rightarrow\pm \infty}|s^*(x)-s_\pm |=0\; ,\;
\lim_{x\rightarrow\pm \infty}r^*(x)=0\; .\; \en

The equations for $\psi$ and
$v$ are rather complicated, but are essentially of the form heat equation
plus irrelevant terms, in the sense of the previous section. Since now
$\psi$
and $v$ decay to zero at infinity, we are in  the situation of that section
and, taking small initial data (in a suitable norm), one shows that the
corresponding solution diffuses to zero. This, then, allows us to prove
equation (4).

\subsection{Stability of Fronts in the Ginzburg-Landau Equation.}

Let us write the Ginzburg-Landau equation (1) in radial and angle
variables,
$ u= re^{i\phi}$:
\be
r_t =r_{xx} + r (1 - \phi^2_x) - r^3 \en
\be
\phi_t = \phi_{xx} + 2 r^{-1} \phi_x  r_x  \en and let us  discuss the
stability of the
front solutions of these equations. It is well known that these equations
have real, positive, front solutions, i.e. solutions of the form
\be
\phi=0, r=r_c (x-ct) \geq 0, \en such that $r_c$ interpolates between a
stable and an unstable solution of (32), i.e. $r_c \rightarrow +1$ for $x
\rightarrow - \infty, r_c \rightarrow 0$, for $x \rightarrow + \infty$.
Indeed, from (32,34), we see that $r_c$ satisfies
\be r_c^{''} + c r'_c + r_c -
r^3_c = 0 \en which, if we reinterpret the variable as ``time", can be seen
as Newton's equation of motion of a particle of mass one subjected to a
friction term $c r'_c$ and to a force deriving from the potential
$\frac{r^2}{2} - \frac{r^4}{4}$, which is an inverted double-well. It is
intuitively clear and easily proved that, for $c$ not too small, solutions
exist that satisfy the required conditions, i.e. such that $r_c$ tends, as
``time" goes to $+ \infty$, to zero, the stable critical point of the
potential, and to one as ``time" goes to $- \infty$. For large ``time"
$u=x-ct,r_c
(u)$ will decay exponentially, as is seen from the
linearization of (35) at
$r=0$. One  gets \be r_c (u) \leq (C_1+C_2u) e^{- \gamma u} \en where
$\gamma$ is given by $\gamma^2 - c\gamma + 1 = 0 $ i.e. \be \gamma_c =
\ha(c
- \sqrt{c^2 -4}) \en which is real for $c \geq 2$, in which case $ \gamma_c
\leq 1$ (actually, one can take $C_2=0$ in (36), if $\gamma <1$).  Thus,
the
larger the friction, the slower the decay. For $c <2$, the solution
``overshoots" the minimum at zero, i.e. $r_c$ is no longer positive. Each
of
the solutions $r_{c}$ with $c \geq 2$ is stable under real perturbations
$(\phi = 0):$  if we start with initial data $r(x,0)$, with $r=r_{c} + s $
with $0 \leq r \leq 1$, $s$ decaying faster than $e^{- \gamma_c x}$ for $x
\rightarrow + \infty$, $r(x,t)$ will converge, as $t \rightarrow + \infty$,
to $r_{c} (x-ct)$, see \cite{AW,Br,CE1}.

However the solution with $c=2, \gamma_c =1$ is more stable than the others
in the sense that any initial data $r (x,0)$ with $0 \leq r \leq 1$ which
decays faster than $e^{-x}$ as $x \rightarrow + \infty$ (in particular, if
$r$ is of compact support) will converge, as $t \rightarrow + \infty$, to
$r_{2} (x-2t)$  \cite{AW,Br}.
 From now on, we shall concentrate on the most interesting front, namely
the
one with $c=2, \gamma = 1$ and we write $r$ for $r_{2}$ .

We consider a complex perturbation of $r : r(x,0) = r + s$ with $\phi
(x,0) \neq 0$ and $\phi (x,0), s(x,0)$ are small in a suitable sense. The
equations satisfied by $\phi$ and $s$ are :
\be
\phi_t = \phi_{xx} + 2(r_x \phi_x + s_x \phi_x) (r +
s)^{-1} \en
\be
s_t = s_{xx} + s (1-3 r^2) - r \phi^2_x
- s \phi^2_x -3 r s^2 - s^3 \en

Since $r$ is a function of $x-2t$ it is convenient to consider also the
equation in the frame of reference of the front : let $u=x-2t$ and $\phi_f
(u,t) = \phi (u+2t,t)$,$s_f (u,t) = s (u+2t,t)$; then $\phi_f$ and
$s_f$ satisfy equations like (38,39), with $2 \phi_{fu}$ added to the RHS
of
(38) and $2 s_{fu}$ added to the one of (39). Now $r = r (u)$ is
time-independent.

To understand the expected behaviour of $\phi_f (u,t)$, let us consider the
linearized equation around the zero solution :
\be
\phi_{ft} =
\phi_{fuu} + 2 r_u \phi_{fu} r^{-1} + 2 \phi_{fu}. \en
 It is convenient to rewrite this equation as a heat equation with a
potential: Let \be \phi_f (u,t) = e^{-u}r(u)^{-1} \psi (u,t).
 \en Then,
$\psi$ satisfies
\be
\psi_t =  \psi_{uu} - V\psi \en with \be V = 1 +
\frac{r''}{r} + 2 \frac{r'}{r}=  r^2. \en
 To derive the last equality, we used eq.(35), satisfied by $r$. Since $r
\simeq
1$ for $u \rightarrow - \infty$, $r \simeq e^{- u}$ for $u \rightarrow +
\infty$, we have $V \simeq 1$ for $u \rightarrow - \infty, V \simeq 0$ for
$u
\rightarrow + \infty$.

So, starting with $\psi (u,0)$ localised around $u=0$,
we expect that
\be \psi (u,t) \sim \left \{ \begin{array}{ll} \frac{ e^{
-t-\frac{u^2}{4t}}}{\sqrt {t}} & u \rightarrow -
\infty \\ \frac{ e^{
-\frac{u^2}{4t} }}{\sqrt t} & u \rightarrow + \infty.
\end{array} \right. \en
Hence, using (41) and the asymptotic behaviour of $r(u)$,
\be
\phi_f (u,t) \sim \left \{ \begin{array}{ll} \frac{e^{-
t-\frac{u^2}{4t}- u}}{\sqrt t} & u \rightarrow - \infty \\ \frac{
e^{-\frac{u^2}{4t}}}{\sqrt t} & u \rightarrow + \infty \end{array} \right.
\en
which can be written as
\be \phi_f (u,t) \sim \left \{ \begin{array}{ll}
\frac{ e^{ - \frac{(u + 2t)^2}{4t} } }{\sqrt t} & u \rightarrow - \infty \\
\frac{ e^{ - \frac{u^2}{4t}}}{\sqrt t} & u \rightarrow + \infty.
\end{array}
\right. \en

Since $u+2t=x$, the first part of (46) is a diffusive wave stationary in
the
fixed frame. The second part represents a diffusive wave which is ``carried
along" by the front. This is a rough, but basically correct picture.

Let us consider the linear equation for $s_f$,  in the front frame :
\be
s_{ft} =  s_{fuu} +  s_{fu} + s_f (1-3 r^2) \en
Writing $s_f=e^{-u}
\sigma$, we get
\be
\sigma_t =  \sigma_{uu} - \tilde{V} \sigma \en
with $\tilde{V} = 3 r^2$.
Following the analysis leading to (44), we get
\be s(u,t) \sim \left \{
\begin{array}{ll} \frac{ e^{-3t-\frac{u^2}{4t}-u} } {\sqrt t} = \frac{
e^{-2t}}{\sqrt t} e^{-\frac{ (u+2t)^2}{4t}} & \mbox{as} \;\;\;  u
\rightarrow
- \infty \\ \frac{ e^{-\frac{u^2}{4t}}} {\sqrt t} e^{-u} & \mbox{as} \;\;\;
 u
\rightarrow + \infty \end{array} \right. \en

There is a ``wave" which is stationary in the front frame, but
exponentially
decreasing in $u$ , while the wave which stays in the fixed frame is
suppressed by the factor $e^{-2t}$.

The rigorous results that justify this picture are of two types:
Using the RG method, Gallay \cite{Ga} was able to obtain very  precise
asymptotics on how a small perturbation of {\it real fronts}  diffuses to
zero (this improves previous results of \cite{Ki,Sa}). So  Gallay considers
equation (39) with $\phi = 0$, in the front frame,  whose linearization is
given by (47). He writes
\be s(u,t) = r' (u) w(u,t) \en
and studies the
behaviour of $w(u,t)$ as $t \to \infty$. Since $r(u)$ goes  exponentially
to
1 or 0 as $u$ goes to $-\infty$ or $+\infty$, $r'(u)$ will  be localized
around $u=0$. The main result is that $w$ has the following  universal
asymptotics \cite{Gal}: let $u=\xi \sqrt{t}$, then  \be w(\xi \sqrt{t},t)
\simeq A t^{-3/2} f^\ast (\xi) \en where  $$ f^\ast(\xi) = \left\{
\begin{array}{lllll} 1 & \mbox{if} \; \xi \leq 0 \\ e^{-\xi^2/4} &
\mbox{if}
\;  \xi>0  \end{array}\right. $$
and $A$ depends again on the initial
conditions. The limit (51) holds in a  weighted $L^1 \cap L^\infty$ norm.
Going back to (47-49), the power  $t^{-3/2}$ is easy to grasp intuitively:
the potential $\tilde V$ in (48) plays  the role of a barrier around $u=0$
($\tilde V$ goes exponentially to 3  for $u \to - \infty$ and exponentially
to $0$ far $u \to + \infty$).  The simplest approximation is to replace the
RHS of (48) by a Laplacian  on ${\bf R}^+$ with Dirichlet boundary
conditions
at $u=0$. And that operator  leads to a $t^{-3/2}$ decay of the solution.
This effect was not taken  into account in (49).

The second type of results deals with the complex perturbations. In
\cite{BK3}, we  consider initial data in the Banach space of
$C^1$-functions
$\phi,s$ with the norm \be \|(\phi,s)\|=\sup_x(1+\mid x
\mid)^{3+\delta}( \mid \phi (x) \mid + \mid  \phi' (x) \mid+(1+e^x)(
\mid s (x) \mid + \mid  s' (x) \mid)) \en and prove

\vs{3mm}

\no {\bf Theorem 3}. {\it For any $\delta > 0$ there exists an $\epsilon >
0$
such that equations $(38,39)$ with   initial data $\phi (x,1) = \phi (x),
s(x,1) = s(x)$, and $\|(\phi,s)\|<\epsilon$,   have a
unique classical
solution $\phi (x,t), s(x,t)$, for all $t \geq 1$, such that} \BE |
\phi (x,t) | & \leq & t^{-\frac{1}{2} + \delta} \\ (1+e^{u}) |s(x,t)| &
\leq & t^{-1+\delta} \EN
\vs{2mm}

\noindent {\bf Remark 1.}
The power laws of the decay in time are presumably
optimal (except for the  $\delta$) and are different from those of Gallay,
because the diffusive  wave that is stationary in the fixed frame, for the
$\phi$ variable (see (45)), goes only diffusively to zero. This in turn
slows
downs
the decay of the  $s$ variable, due to the nonlinear term
$s \phi^2_x$ in (39).

\noindent {\bf Remark 2.} The nonlinear terms in (38,39) turn out to be
irrelevant, in the RG sense.
However, this is not a simple affair: to show it, one has to take into
account  the precise decay of $\phi$ and $s$ both in the fixed and the
front  frames. This makes the proofs rather complicated.

\noindent {\bf Remark 3.} Finally, let us mention that Eckmann and Wayne
\cite{EW}, using a  completely
different (and simpler) method, namely coercive functionals, have proven
similar
results: they can consider a larger space of perturbations $(s,\phi)$
than the one defined by (52), but they do not obtain explicit
 upper bounds on the decay in time.

\section{Universality in Blow-Up for Nonlinear Heat Equations}
\setcounter{equation}{0}

\subsection{Statement of the problem.}

Let us now consider equations for which global existence results do not
hold: the solutions of the initial value problem  \be u_t = u_{xx} + u^p
\en where $p > 1, u=u(x,t),x \in {\bf R}$, and
$u(\cdot,0)=u_0\in C^0({\bf
R})$,  will, for a large class of initial data $u_0$, diverge in a finite
time at a single point (for reviews on this problem, see
\cite{HV,Le}). Again, we
limit ourselves to one space dimension, but the generalization is
straightforward.

The RG ideas can be applied to the analysis of the profile of the solution
at
the time of blow-up. To explain what this means, let us fix the blow-up
point
to be 0 and the blow-up time to be $T$. Then, we ask whether it is possible
to find a function $f^*(x)$ and a rescaling $g(t,T)$ so that \be \lim_{t
\uparrow T} (T-t)^{\frac{1}{p-1}} \; u(g(t,T)x,t)=f^*(x) \en Moreover, we
want to see how $g$ or $f^*$ depend on the initial data.

The prefactor $(T-t)^{\frac{1}{p-1}}$ in (2) can be understood easily:  for
initial data $u_0(x)$ constant in $x$, $u(t)$ solves  the ODE
$ u_t = u^p$,
i.e. $u(t) = ((p-1)(T-t))^{\frac{1}{1-p}}$ for  $T=(p-1)^{-1}u_0^{1-p}$.
 In \cite{HV2,HV3,HV1,V}
(see also \cite{F1,F2}) several possible $f^*$'s are discussed,  and the
set
of initial data that will lead to a given $f^*$ is partially characterized.
In \cite{BK4}, we
 showed that there exists, in the space of initial data $C^0({\bf R})$,
sets
${\cal M}_k$ of codimension $2k$, such that,  for $u_0 \in {\cal M}_k$, the
limiting behaviour (2) is obtained, in the case $k=1$, for  \BE
g(t,T)&=&((T-t) | \log (T-t)|)^{\frac{1}{2}} \\ f^*(x)&=&\left(p-1+
b^*x^2\right)^{\frac{1}{1-p}}  \EN  where $b^*=
 \frac{(p-1)^2}{4p}$, and in the case $k>1$ for \BE
g(t,T)&=&(T-t)^{\frac{1}{2k}}\\ f_b^*(x)&=&\left(p-1+ bx^{2k}
\right)^{\frac{1}{1-p}}. \EN where now $b$ is an arbitary positive number.
As shown in \cite{BK4}, one can also  add suitable (irrelevant) terms to
(1)
without affecting the result. It was shown in \cite{HV2,HV1,V,F2} that,
under
quite general  hypotheses, (3,4) or (5,6) are the only possibilities (see
also \cite{VGH} for a formal analysis). Moreover, solutions that behave as
in
(5,6)  for $k=2$ were constructed in \cite{HV2}.

The codimension of  ${\cal M}_k$ for $k=1$
 is easy to understand: since we have fixed the blow-up  point
 (to zero) and the blow-up time (to $T$), we have to fix two parameters  in
the initial data. To reach the other profiles, $2k-2$ additional parameters
need to be fixed in the initial data. The $k=1$ situation is therefore the
most generic one.

In the RG language,  $f^* $ and $f^*_b$ can be viewed as
 fixed points of a renormalization group transformation having  $2k$
unstable
(``relevant", in renormalization group terminology) directions. Thus, to
converge towards the fixed point,  one has to fine-tune $2k$ parameters
(one
for each unstable direction) and this explains  why ${\cal M}_k$ is of
codimension $2k$, and in what sense $f^*$, $f^*_b$ are ``universal". In
addition, we encounter also one neutral (``marginal") mode, which, for
$k=1$,
turns out to be stable when nonlinear effects are taken into account  and
for
$k>1$ parametrizes a curve of fixed points.

Our results are perturbative, i.e. the sets ${\cal M}_k$  consist of
initial
data that are close to the corresponding fixed point. Therefore, our
results
are similar to those of Bressan
 \cite{Br} who considers a nonlinearity $e^{u}$ instead of $u^p$ and
obtains
the universal profile analogous to our $k=1$ case.  The connection of
blow-up
and center manifold theory was used earlier in the work of Filippas, Kohn
and
Liu \cite{F1,F2} and of Herrero, Velazquez \cite{HV2,HV3,HV1,HV,V} and
 Galaktionov \cite{VGH}. Futhermore, the scaling and the dynamical  systems
aspect of our work goes back to Giga and Kohn \cite{GK,Kohn1,Kohn2}.
Rescalings as in (7) below were used as a technique for numerical
computation
in \cite{Kohn3}.

Let us first describe a change of variables that transforms the problem (1)
into a problem of long time asymptotics: we write (1) in the
``blow--up--variables'':  given a $u: {\bf R} \times [0,T) \to {\bf R}$,
define $\phi :  {\bf R} \times [-\log\ T,\infty) \to {\bf R}$ by \be
u(x,t) = (T - t)^{-{1\over p-1}}\phi ({x\over (T - t)^{1/2k}}, -\log\ (T -
t)). \en Then $u$ is a classical solution of (1) if and only if
 $\phi (\xi,\tau)$ is a classical solution of
\BE
\phi_\tau& =&
L_\tau^{-2} \phi_{\xi\xi} - {1\over 2k}\xi \phi_\xi - {1\over p - 1}\phi +
\phi^p \\ \phi (\xi,\tau_0)& =&  T^{1\over p-1}u_0(T^{1\over 2k}\xi) \EN
where $\tau_0=-\log T$, and \be L_{\tau} = e^{{1\over 2}\tau (1-1/k)}. \en

 We  construct in \cite{BK4} global  solutions of (8), with suitable
initial
data, thereby establishing blow--up for (1).
 Note that, for $k=1$, the scaling in (7) differs from the one used in (3)
by
a factor $| \log (T-t) |^{1/2}=\tau^{1/2}$. Actually,  the situation where
$k>1$ is easier to understand heuristically, so let us start by discussing
the latter.

\subsection{Analysis of $k>1$.}

For $k>1$, as $\tau\to\infty$, the factor $L_{\tau}^{-2}$ in front of the
second derivative in (8) leads us to consider  the solutions of
\be
\phi_\tau = -{1\over 2k}\xi \phi_\xi - {1\over p-1}  \phi + \phi^p\en
Observe that the ''fixed points'' $f_b^*(x)=\left(p-1+ bx^{2k}
\right)^{\frac{1}{1-p}}$ of (6)
 are stationary solutions of that equation.
 The latter can of course be integrated  in closed form, but before doing
that, let us first look at  its linearization around the constant solution
$\phi= (p-1)^{1\over 1-p}$.  The linear problem is $\phi_\tau= {\cal
L}_\infty\phi$,  where
\be
{\cal L}_\infty = - {1\over 2k}\xi \frac{d}{d\xi} + 1 \;
. \en and so, in the space of polynomials,  we have now $2k$ expanding
directions corresponding to $\xi^n$, for $n<2k$.

Equation (11) is solved by putting $\phi (\xi,\tau) = e^{-{\tau\over p-1}}
h(e^{-\tau/2k}\xi,\tau)$ whereby $\frac{d}{d\tau} h(y,\tau) =
e^{-\tau}h(y,\tau)^p$ and so, for $\rho = \tau - \tau_0$ \be \phi
(\xi,\tau) = {e^{-{\rho\over p-1}} f(e^{-\rho/2k}\xi)\over [1 -
(p-1)f(e^{-\rho/2k}\xi)^{p-1}(1 - e^{-\rho})]^{1/p-1}} \en where $\phi
(\xi,\tau_0) = f(\xi)$.
 The stationary solutions  $f_b^*$ are stable in a suitable codimension
$2k$
space: let us consider $f$ smooth, with  \be f(0) = (p - 1)^{-{1\over
p-1}},\
f^{(\ell)}(0) = 0\  \ \ell < 2k,\ f^{(2k)}(0) = \beta < 0\en and \be 0\leq
f(\xi) < (p - 1)^{-{1\over p-1}} \ \;\; \xi \neq 0.\en  Then, for all
$\xi\in{\bf R}$,  \be |\phi (\xi,\tau) - f^*_b(\xi)|
\mathrel{\mathop{\longrightarrow}\limits_{\tau \to
 \infty}} 0\en where  \be f_b^*(\xi) = (p - 1 + b\xi^{2k})^{-{1\over
p-1}}\en
for some
 $b$ depending on $\beta$,$ k$, $p$.

These considerations thus lead us to expect (8) to have  global solutions
with initial data in a suitable
 codimension $2k$ set in a ball around (17) in a suitable Banach space. Of
course, the perturbation $L^{-2}_{\tau}\phi_{\xi\xi}$ in (8) is very
singular, but, basically, the picture is not much modified: the unstable
modes turn out to be $\tau$-dependent
 Hermite functions instead of the monomials $\xi^n$, and one has to
fine-tune
the projection of the initial data on these Hermite functions instead of
imposing the vanishing of the derivatives as in (14). See \cite{BK4} for
details.

\subsection{A non-conventional center manifold problem: $k=1$.}

Consider now the case $k=1$. There are several
 differences with respect to the previous one: there is no damping factor
in
front of the second derivative in (8), and the asymptotics is given by (4),
with a ''universal'' $b^*$, and an extra logarithmic factor in the
definition
(3) of $g(t,T)$. To understand the dynamics of (8) in that case,  let us
start  by considering again its linearization around the constant solution
$\phi=  (p-1)^{1\over 1-p}$. The linear problem
is $ \phi_\tau= {\cal
L}\phi$, where now
\be
  {\cal L} = \frac{d^2}{d\xi^2} - {1\over 2}\xi \frac{d}{d\xi} + 1 \;
. \en   Hence, the first thing we have to do, in order to understand the
stability of the
 constant solution, is to study the spectrum of the linear operator ${\cal
L}$. ${\cal L}$ is self--adjoint on ${\cal D}({\cal L})  \subset L^2({\bf
R},d\mu)$ with \be d\mu (\xi) = \frac{e^{-\xi^2/4}d\xi}{\sqrt{4 \pi}} \en
The
spectrum of ${\cal L}$ is \be spec({\cal L}) = \{1 - {n\over 2}\mid n \in
{\bf N}\} \en and we take as eigenfunctions multiples of Hermite
 polynomials \be h_n(\xi) = \sum\limits_{m=0}^{[{n\over 2}]}{n!\over
m!(n-2m)!}(-1)^m {\xi}^{n-2m} \en that satisfy \be \int h_nh_md\mu = 2^n
n!\delta_{nm} \en and \be {\cal L}h_n = (1 - {n\over 2})h_n. \en Thus the
linearization of (8) at the constant solution,  for $k=1$, has two
expanding (``relevant'') modes, $h_0$ and $h_1$, and one neutral
 (``marginal'') one, $h_{2}=\xi^2-2$.

Our goal is therefore to construct a center manifold for the flow of (8),
in
a neighbourhood of the fixed point. Formally, we would expand,
\be
\phi (\xi,\tau) = (p-1)^{\frac{1}{p-1}} + \psi (\xi,\tau) \;\; \mbox{as}
\;\;
\psi = \sum^\infty_{n=0} \psi_n (\tau) h_n (\xi),
\en
and rewrite
(8) for the $\psi_n(\tau)$ as an infinite set of ODE's:
$$\frac{d}{d\tau} \psi_n (\tau) =
(1-\frac{n}{2}) \psi_n (\tau)  + \; \mbox{nonlinear terms}
$$
A formal solution of this flow yields (see below for the calculation):
$\psi_2 (\tau) \simeq C_p (\log \tau) \tau^{-1}$
with $C_p = \frac{-(p-1)^{\frac{1}{1-p}}}{4p}$ as in (4).

However, there are severe problems with this approach. Since the
eigenfunction $h_n$ of the linearization $\cal L$ {\it increase} at
infinity, the expansion (24) is not useful for $\xi$ large, and, in
particular, we cannot use any standard infinite dimensional center
manifold theorem. The key to the solution to this problem comes again from
a scaling argument, which will explain the emergence of the fixed point
$f^*$. Let \be \phi_L(\xi,\tau) = \phi (L\xi,L^{2}\tau) \en
Then, $\phi_L$
satisfies the equation $$ H(\phi_L) = L^{-2}(- {\phi}_{L\tau} +
\phi_{L\xi\xi}), $$ where we defined
\be
H(\phi)= {1\over 2}\xi \phi_\xi + {1\over p-1}\phi - \phi^p\; . \en Hence,
as
$L \to \infty$,
we expect the solutions of $H(\phi) = 0$ to be relevant.
These are (like the stationary solutions of (11)) given
by the one-parameter family  $f_b^*$, given by (6), with $k=1$. Therefore,
we have the following picture: instead of perturbing around the constant
solution, introduce $
\phi_b(\xi,\tau) = (p-1+ b\xi^{2}/\tau)^{1\over {p-1}}$, and write
 \be \phi (\xi,\tau) = \phi_b(\xi,\tau) + \eta (\xi,\tau). \en
A local (i.e. for $\xi$ small) center manifold analysis then fixes
$b=b^*$, as in (4). For large $\xi$, namely ${\xi^2\over \tau} > {\cal
O}(1)$, the linearization $\tilde{\cal L}$ of (8) around $\phi_b$ differs
from ${\cal L}$,
and, actually, in that region (see below) $\tilde{\cal L} \simeq {\cal
L}-2$.
By (20), the spectrum of ${\cal
L}-2$ is entirely negative. Hence, the dynamics tends to {\it contract}
${\eta}$.

We will now explain the calculation that yields $b^*$ in (4). For
simplicity of notations, we shall consider only $p=2$, i.e. \be
\phi_b(\xi,\tau) = (1+ b\xi^{2}/\tau)^{-1}.\en
 We
get, using $H(\phi_b) = 0$,
\BE
\eta_\tau & =& \eta_{\xi\xi} - H(\phi_b +
\eta) + H(\phi_b)  + \phi_{b\xi\xi} - \phi_{b\tau}    \non\\
&=& ({\cal
L} + W)\eta + M(\eta) + \phi_{b\xi\xi}  -  \phi_{b\tau}   \EN where we
introduce  \BE &&W = 2(\phi_b -1)\\ &&M(\eta) = (\phi_b + \eta)^2 -
\phi_b^2 - 2\phi_b \eta. \EN The operator $\cal L$, given by (18), has
two unstable modes. Note that, formally, (i.e., for $\xi$ of order one) $W$
is ${\cal O}(\tau^{-1}),\ M$ is nonlinear in $\eta$ and
$\phi_{b\xi\xi} -
{\phi}_{b\tau}$ is ${\cal O}(\tau^{-1})$.  We want to construct a
center manifold for (29), i.e.\ to see how to fix two
 parameters in the initial data of $\eta$, such that the flow of (29)
drives
$\eta$ to zero.  A simple calculation will show that this can  only be
achieved through a suitable choice of $b$ in (28).

To see this, let us simplify further and consider
 $\eta$ even in $\xi$, which will imply that we need  to
fix only one
parameter.  This example contains all the relevant features
of the general
case.  It is convenient to write \be \eta_0 (\tau) = \frac{a}{\tau} \en and
define $\psi$ by $$ \eta = \eta_0 + \psi. $$
Then $\psi$ satisfies the
equation
\be
\psi_\tau = \tilde{\cal L} \psi + N(\psi) + \alpha \en
with
\BE
\tilde{\cal L}&=& {\cal L} + V \nonumber\\
V& =& 2(\phi_b + \eta_0 -1)\\ N(\psi) & =& (\phi_b + \eta_0
+ \psi)^2 -  (\phi_b + \eta_0)^2 - 2(\phi_b + \eta_0)\psi \\
 \alpha& = &\phi_{b\xi\xi} -  \phi_{b\tau} +  ({\cal L} + W)\eta_0 -
\eta_{0 \tau}
+ M(\eta_0) \nonumber \\
& =&  \phi_{b\xi\xi}-\phi_{b\tau} + \eta_0+ W
\eta_0 - \eta_{0\tau} + M (\eta_0). \EN

Let us decompose $\psi$ (which is even in $\eta$) as \be \psi =
\psi_0(\tau)
+ \psi_2(\tau)h_2 + \psi^{\perp} \en where $\psi^{\perp}$ is orthogonal  to
$h_n,\ n \leq 2$ . Next we expand $V$ and $\alpha$ (for $\xi={\cal O}(1))$:
\BE &&V = -{2b\xi^2\over \tau} + {2a\over \tau} +  {\cal O} ({1\over
\tau^2})
\\ &&\alpha =(a-2b)\tau^{-1} + (a+a^2+ (12b^2 -b - 2ab)\xi^2))\tau^{-2} +
{\cal O}(\tau^{-3}). \EN
Inserting (32), (37) in (33) and retaining only the
leading terms in $1/\tau$ and $\psi_i,\ i = 0,2$, we get from
$\psi_{i\tau}
= (2^i i!)^{-1}(h_i,\psi_\tau)$ ($(\cdot,\cdot)$  is the scalar product of
$L^2({\bf R}, d\mu)$ and we use (22)):
\BE
\psi_{0\tau} &=& \psi_0 + (a - 2b)\tau^{-1} +
R_0\\
\psi_{2\tau} &=& \beta \tau^{-1} \psi_2 + (12b^2 -b - 2ab)\tau^{-2} +
R_2 \EN
where $R_0 = {\cal O}(\tau^{-2} + \tau^{-1}|\psi| +  |\psi|^2),  R_2
= {\cal O}(\tau^{-3} + \tau^{-1}|\psi_0| +  \tau^{-2}|\psi_2| +
|\psi|^2)$,and
$\beta = 2a - {1\over 4}b(\xi^2h_2,h_2) = 2a - 20b$
 (coming from the $V \psi$ term in (33)). We choose now $a$ so that the
${\cal O}(\tau^{-1})$  term in $\psi_{0 \tau}$ vanishes i.e. \be a = 2b \en
and $b$ such that the ${\cal O}(\tau^{-2})$ term in
 $ {\psi}_{2 \tau}$ is zero: \be b = b^*=1/8. \en
Note that this choice
correspond to $b=b^*$ in (4) for $p=2$. Then $\beta=-2$ and our equations
read
\be
\psi_{0\tau} = \psi_0 + R_0,\ \ \psi_{2 \tau}
 = -{2\over \tau}\psi_2 + R_2. \en Now, keeping in mind the presence of the
$R_0$, $R_2$ terms, \be \psi_0 = {\cal O}(\tau^{-2}),\ \ \psi_2 = {\cal O}
((\log\ \tau)\tau^{-2}) \en would be consistent solutions. Of course, we
need
to show that the  expanding variable $\psi_0$ will satisfy (45) by a
suitable
choice of $\psi_0(\tau_0)$. This is rather easy to do, using the fact that
$\psi_0$ is
 expanding; in the general case (with more than one parameter to fix), we
used a topological argument.

In the rigorous proof, we set up a suitable Banach space for the function
$\eta$, parametrized by the $\psi_i$'s for $\xi^2/\tau<{\cal O}(1)$, and a
function $\eta_l$ for  $\xi^2/\tau>{\cal O}(1)$. The function $\eta_l$ will

contract under the action of $\tilde{\cal L}$: indeed,
from (34,28,32), we see that the
potential $V$ tends to $-2$ as $\xi^2/\tau$ (and $\tau$) $\rightarrow
\infty$.

\section{Open Problems}
\setcounter{equation}{0}
There are several patterns and fronts in dissipative equations,
and their stability can probably be studied using RG methods.
For example, the Cahn-Hillard equation \cite{CH}:
\be
u_t=\Delta (-\Delta u -u -u^3)
\en
is often used to study the phase separation in alloys and fluids.
One would like to study the stability
(and possibly the dynamics), in infinite volume,
of interface solutions of that equation.

Another major open problem consists in the
extension of the RG method to hyperbolic equations. The latter have
their own scaling laws, see \cite{Le,St}.
A lot is known on the
 stability of soliton solutions of (generalized) KdV equations
\cite{PW}, but the
stability of localized solutions of other
nonparabolic equations is quite open.
Also, there are open problem concerning the
 blow-up of solutions, most
notably in the nonlinear Schr\"odinger equation \cite{LPSS}.

So far, we have only considered equations
whose solutions have a rather
simple asymptotic behaviour. Of course, it is well
known that finite
dimensional dynamical systems described by
differential equations can have a
chaotic asymptotic behaviour, i.e. depend sensitively on
the initial data but
have also some statistical regularity in the sense that
the long time average
along the orbits is described by an invariant (SRB) measure
\cite{ER}. For
certain classes of $F$ in (1.1),
one would like to find natural invariant measures for the flow.

A class of dynamical systems, modeling PDE's,
is obtained by discretizing space and time and
considering a recursion \be
u(x,t+1) = F(x,u(\cdot,t)) \en i.e. $u(x,t+1)$, with
$x$ being a site of a
lattice, is determined by the values taken by $u$ at
time $t$ (usually on the
sites in a neighbourhood of $x$). For a suitable class
of $F$'s such
dynamical systems are called Coupled Map Lattices (CML)
\cite{Ka,Ka2}.
The study of the invariant measures, and their properties,
even  for these
"toy models" is essentially a terra incognita. In \cite
{BS,PS,BK6} various
''high-temperatures'' (weak coupling) results are obtained.
Almost nothing is
known about the ''low temperatures'' (strong coupling)
side (see \cite{Bu,BC,MH}). One would like
 to find models for which the set of
invariant measures changes as the coupling is varied; this
phenomenon  would then
be interpreted as a kind of nonequilibrium phase transition.

\vs{5mm}

\no{\Large\bf Acknowledgments}

\vs{3mm}

We would like to thank T. Gallay,
N. Goldenfeld, Y. Oono, J. Taskinen, G. Wayne, and
J. Xin for interesting discussions during
the preparation of this review.
This work was supported by NSF
grant DMS-9205296, by
EC grants SC1-CT91-0695 and CHRX-CT93-0411 and was done in part
during a visit of the authors to the Mittag-Leffler Institute.

\vs{15mm}

\end{document}